\documentclass{ws-procs9x6-cpt16}
\begin{document}

\newcommand{\refeq}[1]{(\ref{#1})}
\def\etal {{\it et al.}}

\def\al{\alpha}
\def\be{\beta}
\def\ga{\gamma}
\def\de{\delta}
\def\ep{\epsilon}
\def\et{\eta}
\def\th{\theta}
\def\ka{\kappa}
\def\la{\lambda}
\def\rh{\rho}
\def\si{\sigma}
\def\ta{\tau}
\def\ph{\phi}
\def\ch{\chi}
\def\ps{\psi}
\def\om{\omega}
\def\Re{\hbox{Re}\,}
\def\Im{\hbox{Im}\,}
\def\re{{\rm Re}}
\def\im{{\rm Im}}
\def\pt#1{\phantom{#1}}
\def\ol#1{\overline{#1}}
\def\kt{{\tilde\ka}}
\def\ktrace{{\kt_{\rm tr}}}
\def\kjm#1#2#3{k^{(#1)}_{(#2)#3}}
\def\cjm#1#2#3{c^{(#1)}_{(#2)#3}}
\def\dit#1{{\pt{---}\mbox{"}\pt{---}}}
\def\ol#1{\overline{#1}}

\title{Limits on Spherical Coefficients in the \\
Minimal-SME Photon Sector }

\author{W.J.\ Jessup and N.E.\ Russell}

\address{Physics Department, Northern Michigan University,
Marquette, MI 49855, USA}

\begin{abstract}
We place limits on spherical coefficients for Lorentz violation
involving operators of dimension four 
in the photon sector of the minimal Standard-Model Extension.
The bounds are deduced from existing experimental results with optical-cavity oscillators.
\end{abstract}

\bodymatter

\section{Introduction}
Coefficients for Lorentz violation 
in the Standard-Model Extension, or SME,\cite{dcak,akgravity} 
have been measured experimentally across the spectrum of 
physics.\cite{datatables} 
A standard Sun-centered inertial reference frame
with orthogonal spatial coordinates $(X,Y,Z)$,
and time $T$,
is used for reporting limits on SME coefficients.\cite{spacepr,2002akmm} 
Most limits on coefficients in the minimal SME, 
involving operators of mass dimension $d=3,4$,
have been expressed using the rectangular $X,Y,Z$ components. 
However, 
for tests of rotational symmetry with higher mass dimensions, 
a spherical decomposition 
of the operators is particularly useful.
In the photon sector of the SME, 
the spherical operators at all mass dimensions
have been calculated.\cite{2009akmm}
Our goal here is 
to obtain limits on the $d=4$ spherical coefficients
from published photon-sector limits expressed in rectangular coordinates.

%%%			------------- CONVERSION TABLE ------------							%%%
\begin{table}										
\tbl{Spherical coefficients for $d=4$ operators in the photon sector}										
{\begin{tabular}{@{}rl@{}}										
\toprule										
	\multicolumn{1}{c}{Spherical}	&	\multicolumn{1}{l}{Rectangular}						\\	
\colrule										
$	\kjm 4 E {20}	$&$	-\sqrt{\frac {6\pi}{5}} (\kt_{e+})^{ZZ}	$					\\ [4 pt]	
$	\Re\kjm 4 E {21}	$&$	\sqrt{\frac{4\pi}{5}}(\kt_{e+})^{XZ}	$					\\ [4 pt]	
$	\Im\kjm 4 E {21}	$&$	-\sqrt{\frac{4\pi}{5}}(\kt_{e+})^{YZ}	$					\\ [4 pt]	
$	\Re\kjm 4 E {22}	$&$	-\sqrt{\frac{\pi} 5}\left\{(\kt_{e+})^{XX}-(\kt_{e+})^{YY}\right\}	$					\\ [4 pt]	
$	\Im\kjm 4 E {22}	$&$	\sqrt{\frac{4\pi} 5}(\kt_{e+})^{XY}	$					\\ [4 pt]	
\colrule										
$	\kjm 4 B {20}	$&$	-\sqrt{\frac {6\pi}{5}} (\kt_{o-})^{ZZ}	$					\\ [4 pt]	
$	\Re\kjm 4 B {21}	$&$	\sqrt{\frac{4\pi}{5}}(\kt_{o-})^{XZ}	$					\\ [4 pt]	
$	\Im\kjm 4 B {21}	$&$	-\sqrt{\frac{4\pi}{5}}(\kt_{o-})^{YZ}	$					\\ [4 pt]	
$	\Re\kjm 4 B {22}	$&$	-\sqrt{\frac{\pi} 5}\left\{(\kt_{o-})^{XX}-(\kt_{o-})^{YY}\right\}	$					\\ [4 pt]	
$	\Im\kjm 4 B {22}	$&$	\sqrt{\frac{4\pi} 5}(\kt_{o-})^{XY}	$					\\ [4 pt]	
\colrule										
$	\cjm 4 I {00}	$&$	\sqrt{4\pi} \ktrace	$					\\ [4 pt]	
\colrule										
$	\cjm 4 I {10}	$&$	-\sqrt{\frac {\pi}{3}} (\kt_{o+})^{XY}	$					\\ [4 pt]	
$	\Re\cjm 4 I {11}	$&$	\sqrt{\frac{\pi}{6}}(\kt_{o+})^{YZ}	$					\\ [4 pt]	
$	\Im\cjm 4 I {11}	$&$	\sqrt{\frac{\pi}{6}}(\kt_{o+})^{XZ}	$					\\ [4 pt]	
\colrule										
$	\cjm 4 I {20}	$&$	-\sqrt{\frac {\pi}{5}} (\kt_{e-})^{ZZ}	$					\\ [4 pt]	
$	\Re\cjm 4 I {21}	$&$	\sqrt{\frac{2\pi}{15}}(\kt_{e-})^{XZ}	$					\\ [4 pt]	
$	\Im\cjm 4 I {21}	$&$	-\sqrt{\frac{2\pi}{15}}(\kt_{e-})^{YZ}	$					\\ [4 pt]	
$	\Re\cjm 4 I {22}	$&$	-\sqrt{\frac{\pi}{30}}\left\{(\kt_{e-})^{XX}-(\kt_{e-})^{YY}\right\}	$					\\ [4 pt]	
$	\Im\cjm 4 I {22}	$&$	\sqrt{\frac{2\pi}{15}}(\kt_{e-})^{XY}	$					\\ [4 pt]	
\botrule										
\end{tabular}										
}										
\label{convert_table}										
\end{table}										
%%%										%%%

\section{Spherical coefficients}
There are 19 independent coefficients for Lorentz violation 
at dimension $d=4$
in the photon sector of the flat-spacetime SME.
In the rectangular basis, 
the usual ones are denoted by 
$\kt_{e+}$,
$\kt_{e-}$, 
$\kt_{o-}$,
each with five components,
$\kt_{o+}$,
with three components,
and 
$\kt_{\rm tr}$.\cite{2002akmm}
In the spherical basis, 
nine of the independent coefficients
are denoted $\cjm 4 I {jm}$,
where $j=0,1,2$ and $|m|=0,1,\ldots j$, 
as is conventional for angular momentum operators.
The remaining ten are
$\kjm 4 E {jm}$ and 
$\kjm 4 B {jm}$,
for $j=2$.
Only a handful of limits on $d=4$ spherical components $\cjm 4 I {2m}$ exist.\cite{2011Parker}
Below, we improve on these and deduce a number of others 
by translation of existing results in the rectangular basis.

To translate limits from one basis to the other, 
the first step is to express the spherical coefficients 
in terms of the rectangular ones. 
One may start with the definitions in Eq.\ (7) of Ref.\ \refcite{2002akmm},
and expand in terms of the spherical coefficients using Table VII of Ref.\ \refcite{2009akmm}.
After applying Eq.\ (45) of Ref.~\refcite{2009akmm} 
to the complex-valued spherical coefficients,
several simplifications occur. 
The results are given in Table \ref{convert_table},
and show a straightforward mapping between the two sets.
In each line of the table,
an equal sign is understood between the spherical coefficient in the first column
and the combination of rectangular ones in the second.
Note that, for $m>0$,
the expressions involve real and imaginary parts of the spherical coefficients.
The simplicity of the results is striking, 
given the expectation that they might involve linear combinations of up to 19 coefficients.

%%%			------ LIMITS ------------						%%%
\begin{table}									
\tbl{Laboratory limits on spherical coefficients 
for $d=4$ operators in the photon sector}									
{\begin{tabular}{@{}rrll@{}}									
\toprule									
	Coefficient	&	\multicolumn{1}{c}{Limit}	&	\multicolumn{1}{c}{System}	&	Reference	\\	
\colrule									
$	\cjm 4 I {10}	$&$	(-7 \pm 10)\times10^{-15}	$&	Optical ring cavity	&	Ref.\ \refcite{2013Michimura}	\\ [4 pt]	
$\phantom{	\cjm 4 I {10}	}$&$	(31 \pm 35)\times10^{-15}	$&	Sapphire cavity oscillators	&	Ref.\ \refcite{2014Nagel}	\\ [4 pt]	
$	\Re\cjm 4 I {11}	$&$	(4 \pm 7)\times10^{-15}	$&	Optical ring cavity	&	Ref.\ \refcite{2013Michimura}	\\ [4 pt]	
$\phantom{	\Re\cjm 4 I {11}	}$&$	(-14 \pm 12)\times10^{-15}	$&	Sapphire cavity oscillators	&	Ref.\ \refcite{2014Nagel}	\\ [4 pt]	
$	\Im\cjm 4 I {11}	$&$	(-4 \pm 9)\times10^{-15}	$&	Optical ring cavity	&	Ref.\ \refcite{2013Michimura}	\\ [4 pt]	
$\phantom{	\Im\cjm 4 I {11}	}$&$	(2 \pm 12)\times10^{-15}	$&	Sapphire cavity oscillators	&	Ref.\ \refcite{2014Nagel}	\\ [4 pt]	
\colrule									
$	\cjm 4 I {20}	$&$	(-3.8\pm 0.9)\times10^{-17}	$&	Rotating optical oscillators	&	Ref.\ \refcite{2016Schiller}	\\ [4 pt]	
$	\Re\cjm 4 I {21}	$&$	(-3.6\pm 2.6)\times10^{-18}	$&	Sapphire cavity oscillators	&	Ref.\ \refcite{2014Nagel}	\\ [4 pt]	
$	\Im\cjm 4 I {21}	$&$	(1.2\pm 2.1)\times10^{-18}	$&\dit{	Sapphire cavity oscillators	}&	Ref.\ \refcite{2014Nagel}	\\ [4 pt]	
$	\Re\cjm 4 I {22}	$&$	(5\pm 11)\times10^{-19}	$&\dit{	Sapphire cavity oscillators	}&	Ref.\ \refcite{2014Nagel}	\\ [4 pt]	
$	\Im\cjm 4 I {22}	$&$	(-5\pm 10)\times10^{-19}	$&\dit{	Sapphire cavity oscillators	}&	Ref.\ \refcite{2014Nagel}	\\ [4 pt]	
\botrule									
\end{tabular}									
}									
\label{limitstable}									
\end{table}									
%%%									%%%

\section{Limits}
Table \ref{convert_table}
shows that each limit on a $d=4$ spherical coefficient
follows from multiplication of the corresponding rectangular limit
by a numerical factor.
In the first ten lines, 
the spherical coefficients
$\kjm 4 E {jm}$ and 
$\kjm 4 B {jm}$
control birefringent effects 
and are constrained at extraordinary levels
by astrophysical observations.\cite{2001akmm}

The nine
$\cjm 4 I {jm}$
coefficients
include $\cjm 4 I {00}=\sqrt{4\pi}\kt_{\rm tr}$,
which has been studied in a variety of contexts 
and limited at multiple levels in numerous ways.\cite{datatables} 
We focus on the eight remaining 
$\cjm 4 I {jm}$ spherical coefficients,
which have been measured in various experiments 
with cavity oscillators.\cite{cavityexpts,2013Michimura,2014Nagel,2016Schiller}

The limits on these spherical coefficients
are presented in Table \ref{limitstable},
where the first column lists the spherical coefficient, 
the second column gives the translated limit, 
and the last two provide a brief description of the system 
and give the reference.
The top portion of the table gives results for $j=1$.
Two results are listed for 
$\cjm 4 I {10}$, 
$\Re \cjm 4 I {11}$, and  
$\Im \cjm 4 I {11}$,
since the relevant experiments have comparable 
sensitivities.\cite{2013Michimura,2014Nagel}
The lower portion of Table \ref{limitstable}
gives limits on the five $j=2$ coefficients
based on recent empirical limits.\cite{2016Schiller,2014Nagel}

As experiments place bounds on spherical SME coefficients,
these results provide a comparison point for photon-sector results that predate 
the development of the spherical formalism. 
We also note that conversions of this type are possible in other sectors of the SME, 
and for higher mass-dimension operators.

\end{document}